\batchmode
\makeatletter
\def\input@path{{C:/Users/Achel/UAI/DISC/Research/Stability//}}
\makeatother
\documentclass[english]{article}
\usepackage{lmodern}
\usepackage[T1]{fontenc}
\usepackage[latin9]{inputenc}
\usepackage{geometry}
\usepackage{textcomp}
\usepackage{amsthm}
\usepackage{amsmath}
\usepackage{graphicx}

\makeatletter
\numberwithin{equation}{section}
\numberwithin{figure}{section}
\newcommand{\lyxaddress}[1]{
\par {\raggedright #1
\vspace{1.4em}
\noindent\par}
}

\usepackage[sort&compress,numbers]{natbib}
\usepackage{hyperref}

\@ifundefined{showcaptionsetup}{}{%
 \PassOptionsToPackage{caption=false}{subfig}}
\usepackage{subfig}
\makeatother

\usepackage{babel}
\begin{document}

\title{\textbf{\Large{}Dynamics and Stability in Retail Competition}}

\author{\noindent Marcelo J. Villena and Axel A. Araneda\thanks{Email: axel.araneda@uai.cl, Tel.: +56 2 23311416}}

\maketitle

\lyxaddress{\begin{center}
Faculty of Engineering \& Sciences, Universidad Adolfo Ib\'a\~nez\\
Avda. Diagonal las Torres 2640, Peñalol\'en, 7941169, Santiago, Chile.
\par\end{center}}

\thispagestyle{empty}

Retail competition today can be described by three main features:
i) oligopolistic competition, ii) multi-store settings, and iii) the
presence of large economies of scale. In these markets, firms usually
apply a centralized decisions making process in order to take full
advantage of economies of scales, e.g. retail distribution centers.
In this paper, we model and analyze the stability and chaos of retail
competition considering all these issues. In particular, a dynamic
multi-market Cournot-Nash equilibrium with global economies and diseconomies
of scale model is developed. We confirm the non-intuitive hypothesis
that retail multi-store competition is more unstable that traditional
small business that cover the same demand. The main sources of stability
are the scale parameter and the number of markets. \textbf{}\\

\textbf{Keywords}: Multi-market Oligopoly, Cournot-Nash competition,
Economies of Scale, Stability, Bifurcations, and Chaos.\\

\section{Introduction}

In an oligopolistic setting under a Cournot scheme \cite{cournot1838recherches},
the strategy of each economic player depends on its own quantity decision,
and on its rival's reaction. Puu was the first to explicitly show
the complex dynamics of the oligopolistic setting under simple assumptions
(isoelastic demand function and constant marginal cost) for two and
three players \cite{puu1991chaos,puu1996complex,puu1998chaotic}.
This kind of analysis has grown significantly during the last decade
in both, the mathematics and complex systems literature, as well as
in the economically-oriented journals.

Indeed, since the Puu's approach, several games has been developed
for the study of the market stability, focusing on: different demand
or price function \cite{askar2015dynamic,Naimzada2006707}, number
of players \cite{ahmed1998dynamics,puu2008stability}, behavioral
assumptions (naive \cite{chen2007controlling,matsumoto2006controlling,bischi2007oligopoly},
versus adaptive \cite{bischi2001equilibrium,bischi2012routes}, bounded
rationality \cite{agiza2002complex,Naimzada2006707,yali2011dynamics}
or heterogeneous expectations \cite{agiza2003nonlinear,agiza2004chaotic,angelini2009bifurcation,ma2011complex}).
In terms of the cost function definition, several developments has
been proposed as well, as non-linear cost function \cite{yali2011dynamics,dubiel2011nonlinear,Naimzada2006707,ma2011complex,bischi2012routes},
capacity constraints |\cite{puu2003cournot,puu2006dynamics,bischi2012routes,lamantia2011nonlinear}
and some spillover effects \cite{bischi2002nonlinear,bischi2002chaos,Bischi2012,bischi2012dynamic}. 

Most works in this line of research have concentrated in single markets
with linear production structures (i.e. assuming constant returns
to scale). Nevertheless, oligopolistic competition today seems to
present multi-market phenomena and, in some cases, they showcase important
economies of scale, especially in the retail industry. Indeed, supermarket
chains and retailers of food, gasoline, supplies and services all
compete for market share through multi-store formats over geographically
separated markets. This localized competition is presented in different
levels: city, region, or country. In this context, companies segment
their strategies, tailoring their selected outcome for different types
of consumers and competitors, which vary by geographical location.
On the other hand, on the supply side, multi-market retailers usually
try to take full advantages of their size, in other words, their economies
of scale. For instance, through the development of distributions centers
that attend most of the stores in an specific territory. Thus, as
the cost structure of multi-market retails depends on the total volume
of the produced goods, the individual cost structure of each store
is usually coupled with the whole business. It is important to point
out that, this system of production, implies a centralized decisions
making process, which become in practice an extremely difficult task.
Summing up, retail competition today can be described by three main
features: i) oligopolistic competition, ii) multi-store setting, iii)
the presence of economies of scale.

Applications of the Cournot scheme into the multi-market problem has
been proposed before by economists, for example, in the case of international
trade. Some of these works modeled the presence of economies of scale,
for the domestic and foreign markets, considering the size (quantity
produced) and other properties of firms \cite{brander1981intra,brander1983reciprocal,Krugman1984,Dixit1984,helpman1984increasing,ben1988oligopoly,leahy1991import,murray1999universal}.
Thus, for instance in a work of Krugman \cite{Krugman1984}, a multi-market
Cournot model with economies of scale was used to explain the successful
performance of Japan as an exporting country at the beginning of the
1980s.

In theoretical terms, the multi-market oligopoly framework was revisited
and generalized in the seminal paper by Bulow, Geanakoplos and Klemperer
\cite{bulow1985multimarket}. One of their main remarks is that the
presence of a multi-store firm in a market may affect the position
in the others for the presence of demand and/or supply spillovers.
In the same line, Bernheim and Whinston, \cite{bernheim1990multimarket},
show that with scale economies, the multi-market contact may produce
\textquotedblleft spheres of influence\textquotedblright{} \cite{edwards1955conglomerate},
that occurs when each of the multi-market competing firms may be more
efficient in some subset of these markets and less efficient in others
(symmetric advantage) or when one firm is more efficient in all markets
(absolute advantage). Despite these multi-market analysis, this literature
has focused mainly on the demand side of the problem, not the supply
side. Specifically, they refer to multi-market contact, when demands
curves recognize substitution and complementarity of different products.

In terms of the analysis of the dynamics of the multi-market Cournot
problem, we found only a few papers \cite{wu2014chaos,ahmed2014controls,andaluz2015dynamics,naimzada2012dynamic},
focusing on different products and scope.

In this context, this research deals with the analysis of stability
and chaos of multi-market competition in the presence of economies
and diseconomies of scale, extending this way the analysis of the
dynamics of the oligopolistic competition. Thus, we model the main
characteristics of the retail competition today, analysing the dynamics
and stability of this particular dynamic system, and we compare these
results with the stability analysis of traditional small business
that cover the same demand, the classic Pu's formulation. 

The main hypothesis of the paper is that non-linear cost structures
in multi-market setting are important sources of instability in the
game outcome. Particularly, we study the stability of a multi-market
Cournot-Nash equilibrium with global economies of scale, that is,
the scale level that is related to the total production of firms,
in all markets, as opposed to local economies of scale presented at
each store individually or linear production structures. In this setting,
the internal organization of a firm may affect its performances over
the markets and the global equilibrium \cite{barcena1999should}.
For example, multi-market firms that buy their products in a centralized
manner, storing them in a distribution center, to be redistributed
afterwards to their retailers store in all markets usually operate
this way to obtain economies of scale in the process of buying and
distribution. In this paper, we assume this type of centralized structure
where companies takes advantage of their size, under economies of
scale, that allow them to decrease the cost structure \cite{qian1993china}. 

This papers is organized as follows: in section 2 classical models
of the Cournot problem are described and extended to the multi-market
framework. In the section 3 a Multi-market-Cournot problem is presented,
considering interrelated cost structures and economies of scale. In
section 4, the study of the stability of the system is addressed and
generalized for the duopoly case. In the fifth section, the complex
dynamics of the game for different numbers markets and values of the
scale parameter are shown by path graphics and bifurcation diagrams.
Finally, the main conclusions for this work are presented.\\

\section{Baseline: Single market oligopoly models\label{Fisher} }

The well-known Cournot-Theocharis model \cite{Theocharis1959,palander1936instability,puu2006rational,canovas2008cournot}
proposed a Cournot Oligopoly model with inverse lineal demand function
and constant marginal cost, that is:

\begin{equation}
P=a-\sum q^{i}
\end{equation}

\begin{equation}
C^{i}=c^{i}q^{i}\label{eq:costo total}
\end{equation}
\\

\noindent where $i=1,\ldots,N$ identify the player, $P$ is the
price of the good, $C$ is the total cost and $q$ is the quantity.

The profit is obtained subtracting the revenues by total cost:

\[
\pi^{i}=q^{i}P\left(q^{1},\ldots,q^{N}\right)-c^{i}q^{i}
\]
\\

The optimization problem (maximization of profit) arrives to:

\begin{equation}
\stackrel{*}{q^{i}}=\frac{a-c^{i}}{2}-\sum_{k\neq i}\frac{q^{k}}{2}\label{eq:theo}
\end{equation}
\\

With solution (Cournot-Nash equilibrium point):

\[
\stackrel{*}{q^{i}}=\frac{a-c^{i}+N\left(\bar{c}-c^{i}\right)}{N+1}
\]
\\

\noindent with $\bar{c}=\left(1/N\right)\sum_{i}c^{i}$

In order to transform the static game \ref{eq:theo} into a dynamic
one, the Cournot or naive strategy is used (see Cournot \cite{cournot1838recherches}
and Puu \cite{puu1991chaos})\footnote{In this strategy, it is assumed that a player at time of the decision
making process, it takes into account the rival's previous output,
e.g, managers use past information and supposes it doesn't change
at this moment.}. Thus, the long run map as an iterative process is given by:\\
\begin{equation}
q^{i}(t+1)=\frac{a-c^{i}}{2}-\sum_{k\neq i}\frac{q^{k}(t)}{2}\label{eq:theoRF}
\end{equation}
\\

The dynamical system of $N$ reaction functions defined by \ref{eq:theoRF}
has a stable equilibrium (fix point) for $n\leq2$ . For $n=3$ the
equilibrium is neutrally stable (stationary oscillations) and for
$n\geq4$ the equilibrium becomes unstable.

A generalization of the Cournot-Theocharis problem was developed by
Fisher \cite{fisher1961stability}. This research allows to work with
increasing or decreasing returns to scale (i.e., economies or diseconomies
of scale). In order to get this result, we need to add a non-linear
term to the traditional linear cost function: 

\begin{eqnarray*}
P & = & a-\sum_{i}q^{i}\\
C^{i} & = & c^{i}q^{i}+d\left(q^{i}\right)^{2}
\end{eqnarray*}
\\

Thus, the profit takes the following form:

\begin{equation}
\pi^{i}=q^{i}P\left(q^{1},\ldots,q^{N}\right)-c^{i}q^{i}-d\left(q^{i}\right)^{2}\label{eq:PROFIT}
\end{equation}
\\

Some restrictions for avoiding non-negativity of outputs, price, profits
and marginal cost are considered: $c>0$, $a\geq c$ and $d>-1/2$. 

Thus, the maximization of the profit (eq. \ref{eq:PROFIT}) leads
to the best response of the $i^{th}$- firm:

\begin{equation}
\stackrel{*}{q^{i}}=\frac{a-{\displaystyle \sum_{k\neq i}q^{k}}-c^{i}}{2\left(1+d\right)}\label{eq:fisher-1}
\end{equation}
\\

Then, the Cournot-Nash equilibrium is given by:

\[
\frac{\left(a-c^{i}\right)\left(1+2d\right)+N\left(\bar{c}-c^{i}\right)}{4d^{2}+2\left(N+2\right)d+N+1}
\]
\\

Hence, the naive dynamics takes the form:

\begin{equation}
q^{i}(t+1)=\frac{a-{\displaystyle \sum_{k\neq i}q^{k}(t)}-c^{i}}{2\left(1+d\right)}\label{eq:fisher}
\end{equation}
\\

Finally, this dynamical system becomes stable when $\left(N-3\right)/2<d$
. Thus, if there are two players the game has a stable equilibrium
if $d>-1/2.$\\

The approaches revised above (Teocharis and Fisher), were design to
model a single-market oligopoly problem (for example the rivalry between
``mom-and-pop'' stores). In this case, both the prices and the costs
don't depends upon the behavior of the players in other markets, because
they did not consider the case of large corporations, with multiple
operations in various locations. 

In the next section, and taking as baseline the models presented above,
we will develop a multi-market Cournot model with economies of scale
that will allow us to describe the modern retail competition.\\

\section{The Model}

Let us consider a multi-market oligopoly where $N$ single-product
firms compete in $M$ markets. All markets are very far away, so there
not exist arbitrage possibilities. Each market has its own price (from
a linear demand function). Then, if $q_{j}^{i}$ is the quantity of
the $i^{th}$ company at the market $j$ and $Q_{j}$ the market supply,
the selling price at the $j^{th}$ market is given by,

\begin{equation}
P_{j}=a_{j}-\sum_{i}q_{j}^{i}=a_{j}-Q_{j}
\end{equation}

We assume a centralized managerial structure, where the production
costs depends of the total outputs of each firm. Besides, due both
their size and specialization the company can have economies of scale.
So we have:

\begin{equation}
C^{i}=c^{i}\sum_{j}q_{j}^{i}+d\left(\sum_{j}q_{j}^{i}\right)^{2}=c^{i}Q^{i}+d\left(Q^{i}\right)^{2}
\end{equation}

\noindent with $c^{i}$ greater than zero. Depending of the value
of $d$ the companies operates under economies of scale ($d<0$) or
diseconomies of scale ($d>0$). The allowable range for the parameter
$d$ will be analyzed later.

The square form of the production cost was used previously for other
scholars in a single-market context \cite{fisher1961stability,du2009analysis,Naimzada2006707,yao2006complex}.
However, for this multi-market scheme we use the non-linearity for
to couple the costs and to enable the existence of economies (diseconomies)
of scale. This is a realistic approach for the retails firms, because
they produce or buy in large scale. 

Under this cost structure, $c^{i}>0$ and $dQ_{max}^{i}>-c^{i}/2$
, where $Q_{max}^{i}$ is the maximum level of production of firm
$i$. The theoretical maximum production (extreme case) is achieved
when a company becomes a monopoly in all the markets (i.e. $Q_{j}=q_{j}^{i},$$\forall j$).
As $P_{j}\geq0$, we have $\sum_{j}a_{j}>Q_{max}^{i}$, and then we
have $2d>c^{i}/\sum_{j}a_{j}$.

Thus the profit function of each firm depends upon each market price,
the quantity sold by the firm in that particular market, and the total
cost of the firm, that considers the sum of the global cost and the
local cost:

\begin{eqnarray}
\pi^{i} & = & \sum_{j}q_{j}^{i}P_{j}-C^{i}\\
 & = & \sum_{j}q_{j}^{i}\left(a_{j}-b_{j}\sum_{i}q_{j}^{i}\right)-c^{i}Q^{i}-d\left(Q^{i}\right)^{2}
\end{eqnarray}

The managerial decision of each firm is to choose the quantity $q_{j}^{i}$
that maximizes its profits. In other words, the $i^{th}$ company
which produces a total output of $Q^{i}$ divided over the $M$ different
markets according to the output vector $\underline{Q}^{i}=\left\{ q_{1}^{i},q_{2}^{i},\ldots,q_{M}^{i},\right\} $
needs to fix the optimal allocation given by $\stackrel{*}{\underline{Q}^{i}}=\left\{ \stackrel{*}{q_{1}^{i},}\stackrel{*}{q_{2}^{i}},\ldots,\stackrel{*}{q_{M}^{i}}\right\} $,
where each one of the $\stackrel{*}{q_{j}^{i}}$ is a solution of
the $M\times N$ simultaneous equations which comprising the first
order conditions of this game given by:

\begin{center}
\begin{eqnarray}
\frac{\partial\pi^{i}}{\partial q_{j}^{i}} & = & a_{j}-Q_{j}-q_{j}^{i}-c^{i}-2dQ^{i}=0\label{eq:FOC-1}
\end{eqnarray}

\par\end{center}

Defining the residual market supply for $i$ as $\tilde{Q}_{j}^{i}$
= $Q_{j}-q_{j}^{i}$, and the participation of the firm $i$ in other
markets different to $j$ as $\hat{Q}_{j}^{i}$ = $Q^{i}-q_{j}^{i}$,
the equilibrium quantity take the value:\\

\begin{equation}
\stackrel{*}{q_{j}^{i}}=\frac{a_{j}-\tilde{Q}_{j}^{i}-c^{i}-2d\hat{Q}_{j}^{i}}{2\left(1+d\right)}\label{eq:q_eq}
\end{equation}

Clearly, the allocation decision depends on the decision of the other
players on this market, and also depends on the participation of the
firm in other markets (or the total firm supply). This results is
consistent with the Cournot intuition and consistent to previous results
for the single-market problem\footnote{When $\tilde{Q}^{i}=0$ } \cite{fisher1961stability,elabbasy2009analysis,dubiel2011nonlinear}.

In order to keep the game on rails, we have that $\stackrel{*}{q_{j}^{i}}$
is a maximum if and only if $d>-1$. Also the marginal profit for
zero output must be non-negative \cite{fisher1961stability}, so $a_{j}\geq c^{i},\forall i$.
This conditions joined with the previous restrictions arrives to: 

\begin{equation}
d>-\frac{1}{2m}\label{eq:d}
\end{equation}
\\

The profit of each firm at the equilibrium point is given by:

\begin{equation}
\pi^{i}={\displaystyle \sum_{j}\left(\stackrel{*}{q_{j}^{i}}\right)^{2}+d\left(\sum_{j}\stackrel{*}{q_{j}^{i}}\right){}^{2}}
\end{equation}

The Jacobian matrix at the equilibrium is given by:

\begin{equation}
J=\begin{bmatrix}{\displaystyle \frac{\partial\stackrel{*}{\underline{Q}^{1}}}{\partial\underline{Q}^{1}}} & {\displaystyle \frac{\partial\stackrel{*}{\underline{Q}^{1}}}{\partial\underline{Q}^{2}}} & \cdots & {\displaystyle \frac{\partial\stackrel{*}{\underline{Q}^{1}}}{\partial\underline{Q}^{N}}}\\
{\displaystyle \frac{\partial\stackrel{*}{\underline{Q}^{2}}}{\partial\underline{Q}^{1}}} & {\displaystyle \frac{\partial\stackrel{*}{\underline{Q}^{2}}}{\partial\underline{Q}^{2}}} & \cdots & {\displaystyle \frac{\partial\stackrel{*}{\underline{Q}^{2}}}{\partial\underline{Q}^{N}}}\\
\vdots & \ddots &  & \vdots\\
{\displaystyle \frac{\partial\stackrel{*}{\underline{Q}^{N}}}{\partial\underline{Q}^{1}}} & {\displaystyle \frac{\partial\stackrel{*}{\underline{Q}^{N}}}{\partial\underline{Q}^{2}}} & \cdots & {\displaystyle \frac{\partial\stackrel{*}{\underline{Q}^{N}}}{\partial\underline{Q}^{N}}}
\end{bmatrix}\label{eq:Jacobian}
\end{equation}
\\
\noindent where each one of the $N\times N$ entries of $J$ is a
$M\times M$ block matrix, that represents the change of the $i$-firm's
reaction functions with respect to the outputs of $j$, with:

\begin{equation}
{\displaystyle \frac{\partial\stackrel{*}{\underline{Q}^{i}}}{\partial\underline{Q}^{k}}}=\begin{bmatrix}{\displaystyle \frac{\partial\overset{*}{q}_{1}^{i}}{\partial q_{1}^{k}}} & {\displaystyle \frac{\partial\overset{*}{q}_{1}^{i}}{\partial q_{2}^{k}}} & \cdots & {\displaystyle \frac{\partial\overset{*}{q}_{1}^{i}}{\partial q_{M}^{k}}}\\
{\displaystyle \frac{\partial\overset{*}{q}_{2}^{i}}{\partial q_{1}^{k}}} & {\displaystyle \frac{\partial\overset{*}{q}_{2}^{i}}{\partial q_{2}^{k}}} & \cdots & {\displaystyle \frac{\partial\overset{*}{q}_{2}^{i}}{\partial q_{M}^{k}}}\\
\vdots & \ddots &  & \vdots\\
{\displaystyle \frac{\partial\overset{*}{q}_{M}^{i}}{\partial q_{1}^{k}}} & {\displaystyle \frac{\partial\overset{*}{q}_{M}^{i}}{\partial q_{2}^{k}}} & \cdots & {\displaystyle \frac{\partial\overset{*}{q}_{M}^{i}}{\partial q_{M}^{k}}}
\end{bmatrix}\label{eq:Jacobian_markets}
\end{equation}

According to our optimal outputs (Eq. \ref{eq:q_eq}), we have:

\begin{equation}
{\displaystyle \frac{\partial\stackrel{*}{\underline{Q}^{i}}}{\partial\underline{Q}^{k}}}=\begin{cases}
-\frac{1}{2\left(1+d\right)}I_{M\times M} & ,\textrm{ for }i\neq j\\
H & ,\textrm{ for }i=j
\end{cases}\label{eq:Jacobian_markets-1}
\end{equation}

\noindent where $I$ is the identity matrix and $H$ is a zero-diagonal
matrix with all the off-diagonal entries equal to $-d/(1+d)$. \\

\section{Dynamic Analysis of the Equilibrium}

For the dynamic modeling, we use naive expectations. Thus, the game
is developed on discrete time as follow\footnote{The dynamic proposed in \ref{eq:qt+1} can lead to negative outputs
without economic sense. Following \cite{agliari2002cournot}, we will
differentiate between two types of trajectories: admissible and feasible.
Calling $T$ the map defined by the $2m$ equations of the game, the
set of admissible ($S$) and feasible points ($F$) is defined respectively
by:

\begin{eqnarray*}
S & = & \left\{ \left(q_{1}^{1},q_{2}^{1},\ldots,q_{M}^{1},q_{1}^{2},q_{2}^{2},\ldots,q_{2}^{M}\right)\in R_{+}^{2M}:T^{n}\left(q_{1}^{1},q_{2}^{1},\ldots,q_{M}^{1},q_{1}^{2},q_{2}^{2},\ldots,q_{2}^{M}\right)\in R^{2M}\quad\forall n>0\right\} \\
F & = & \left\{ \left(q_{1}^{1},q_{2}^{1},\ldots,q_{M}^{1},q_{1}^{2},q_{2}^{2},\ldots,q_{2}^{M}\right)\in R_{+}^{2M}:T^{n}\left(q_{1}^{1},q_{2}^{1},\ldots,q_{M}^{1},q_{1}^{2},q_{2}^{2},\ldots,q_{2}^{M}\right)\in R_{+}^{2M}\quad\forall n>0\right\} 
\end{eqnarray*}

Then, the mathematical results are based on the admissible trajectories.
However, in the economic context, only the feasible points will be
considered.}:

\begin{equation}
q_{j}^{i}(t+1)=\frac{a_{j}-\tilde{Q_{j}^{i}}(t)-c^{i}-2d\hat{Q}^{i}(t)}{2\left(1+d\right)}\label{eq:qt+1}
\end{equation}
\\

When there are two players competing over $M$ different markets,
the long-run map proposed in \ref{eq:qt+1} is defining for the following
$2m$ equations:

\begin{eqnarray}
q_{1}^{1}(t+1) & = & \frac{a_{1}-q_{1}^{2}(t)-c^{1}-2d\left(q_{2}^{1}(t)+q_{3}^{1}(t)+\ldots+q_{M}^{1}(t)\right)}{2\left(1+d\right)}\label{eq:dyn2xm}\\
\vdots &  & \vdots\nonumber \\
q_{j}^{1}(t+1) & = & \frac{a_{j}-q_{j}^{2}(t)-c^{1}-2d\left(q_{1}^{1}(t)+\ldots+q_{j-1}^{1}(t)+q_{j+1}^{1}(t)\ldots+q_{M}^{1}(t)\right)}{2\left(1+d\right)}\nonumber \\
\vdots &  & \vdots\nonumber \\
q_{M}^{1}(t+1) & = & \frac{a_{M}-q_{M}^{2}(t)-c^{1}-2d\left(q_{1}^{1}(t)+q_{3}^{1}(t)+\ldots+q_{M-1}^{1}(t)\right)}{2\left(1+d\right)}\nonumber \\
q_{1}^{2}(t+1) & = & \frac{a_{1}-q_{1}^{1}(t)-c^{2}-2d\left(q_{2}^{2}(t)+q_{3}^{2}(t)+\ldots+q_{M}^{2}(t)\right)}{2\left(1+d\right)}\nonumber \\
\vdots &  & \vdots\nonumber \\
q_{j}^{2}(t+1) & = & \frac{a_{j}-q_{j}^{1}(t)-c^{2}-2d\left(q_{1}^{2}(t)+\ldots+q_{j-1}^{2}(t)+q_{j+1}^{2}(t)\ldots+q_{M}^{2}(t)\right)}{2\left(1+d\right)}\nonumber \\
\vdots &  & \vdots\nonumber \\
q_{M}^{2}(t+1) & = & \frac{a_{M}-q_{M}^{1}(t)-c^{2}-2d\left(q_{1}^{2}(t)+q_{3}^{2}(t)+\ldots+q_{M-1}^{2}(t)\right)}{2\left(1+d\right)}\nonumber 
\end{eqnarray}
\\

The nontrivial Cournot-Nash equilibrium point for the previous set
of equations (static solution) is given by:

\begin{equation}
\stackrel{*}{q_{j}^{i}}=\frac{\left(a_{j}-c^{i}\right)\left(1+2Md\right)+c^{k}-c^{i}}{3+\left(2M\right)^{2}d^{2}+8Md}+\frac{2}{3}\frac{M\left(a_{j}-\bar{a}\right)\left(2Md^{2}+d\right)}{3+\left(2M\right)^{2}d^{2}+8Md},\quad d\neq-\frac{3}{2M},-\frac{1}{2M}\label{eq:cournot point dupoly-1}
\end{equation}
\\

Using the equations\ref{eq:Jacobian}, \ref{eq:Jacobian_markets}
and \ref{eq:Jacobian_markets-1} the Jacobian matrix for the system
is defined by:

\begin{equation}
J_{2xM}=\begin{bmatrix}{\textstyle {\scriptstyle \ensuremath{M}\textrm{times}}}\left\{ \underbrace{\begin{array}{cccc}
0 & -\frac{d}{1+d} & \cdots & -\frac{d}{1+d}\\
-\frac{d}{1+d} & 0 & \cdots & -\frac{d}{1+d}\\
\vdots & \ddots &  & \vdots\\
-\frac{d}{1+d} & -\frac{d}{1+d} & \dots & {\displaystyle 0}
\end{array}}\right. & \begin{array}{cccc}
-\frac{1}{2\left(1+d\right)} & {\displaystyle 0} & \cdots & {\displaystyle 0}\\
{\displaystyle 0} & -\frac{1}{2\left(1+d\right)} & \cdots & {\displaystyle 0}\\
\vdots & \ddots &  & \vdots\\
{\displaystyle 0} & 0 & \cdots & -\frac{1}{2\left(1+d\right)}
\end{array}\\
{\textstyle {\scriptstyle \ensuremath{M}\textrm{times}}}\\
\begin{array}{cccc}
-\frac{1}{2\left(1+d\right)} & 0 & \cdots & 0\\
{\displaystyle 0} & -\frac{1}{2\left(1+d\right)} & \cdots & {\displaystyle 0}\\
\vdots & \ddots &  & \vdots\\
{\displaystyle 0} & 0 & \cdots & -\frac{1}{2\left(1+d\right)}
\end{array} & \begin{array}{cccc}
0 & -\frac{d}{1+d} & \cdots & -\frac{d}{1+d}\\
-\frac{d}{1+d} & 0 & \cdots & -\frac{d}{1+d}\\
\vdots & \ddots &  & \vdots\\
-\frac{d}{1+d} & -\frac{d}{1+d} &  & {\displaystyle 0}
\end{array}
\end{bmatrix}\label{eq:Jacobian-1}
\end{equation}
\\

The characteristic equation of \ref{eq:Jacobian-1} takes the form:

\begin{equation}
{\textstyle \left(\lambda-\frac{1}{2}\frac{2d+1}{\left(1+d\right)}\right)^{m-1}\left(\lambda-\frac{1}{2}\frac{2d-1}{\left(1+d\right)}\right)^{m-1}\left(\lambda+\frac{1}{2}\frac{2\left(m-1\right)d+1}{\left(1+d\right)}\right)\left(\lambda+\frac{1}{2}\frac{2\left(m-1\right)d-1}{\left(1+d\right)}\right)=0}\label{eq:roots}
\end{equation}
\\

Thus, we have four\footnote{If we fix $M=1$, we have only the two first eigenvalues.}
different eigenvalues:

\begin{eqnarray}
\lambda_{1} & = & -\frac{2\left(m-1\right)d+1}{2\left(1+d\right)}\label{eq:eigenvalues}\\
\lambda_{2} & = & -\frac{2\left(m-1\right)d-1}{2\left(1+d\right)}\nonumber \\
\lambda_{3,\ldots,m+1} & = & \frac{2d+1}{2\left(1+d\right)}\nonumber \\
\lambda_{m+2,\ldots,2m} & = & \frac{2d-1}{2\left(1+d\right)}\nonumber 
\end{eqnarray}

The local stability of equilibrium is achieved only if each eigenvalue
is within the unit circle. According to \ref{eq:eigenvalues} this
is fulfilled when:

\begin{equation}
{\displaystyle \begin{cases}
{\displaystyle -\frac{1}{2m}}<d<{\displaystyle \frac{1}{2\left(m-2\right)}}, & \quad m\neq2\\
{\displaystyle d>-\frac{1}{4}}, & \quad m=2
\end{cases}}\label{eq:results duopoly}
\end{equation}
\\
By \ref{eq:results duopoly} is clear that stability of the duopoly
game depends on both the scale parameter $(d)$ and the number of
markets $(m)$. In the fig. \ref{fig:Stabililty-zone} the relationship
between $m$ and $d$, in order of to arrive to the stability of the
game is shown. When m=1,2 the equilibrium becomes unstable only if
d>-1/(2m). However, if m>3, we need to put an upper bound for the
stability condition. We see that when the number of markets increases,
the stability is achieved only if d tends to zero.\\

\begin{center}
\begin{figure}
\begin{centering}
\includegraphics[width=0.6\textwidth]{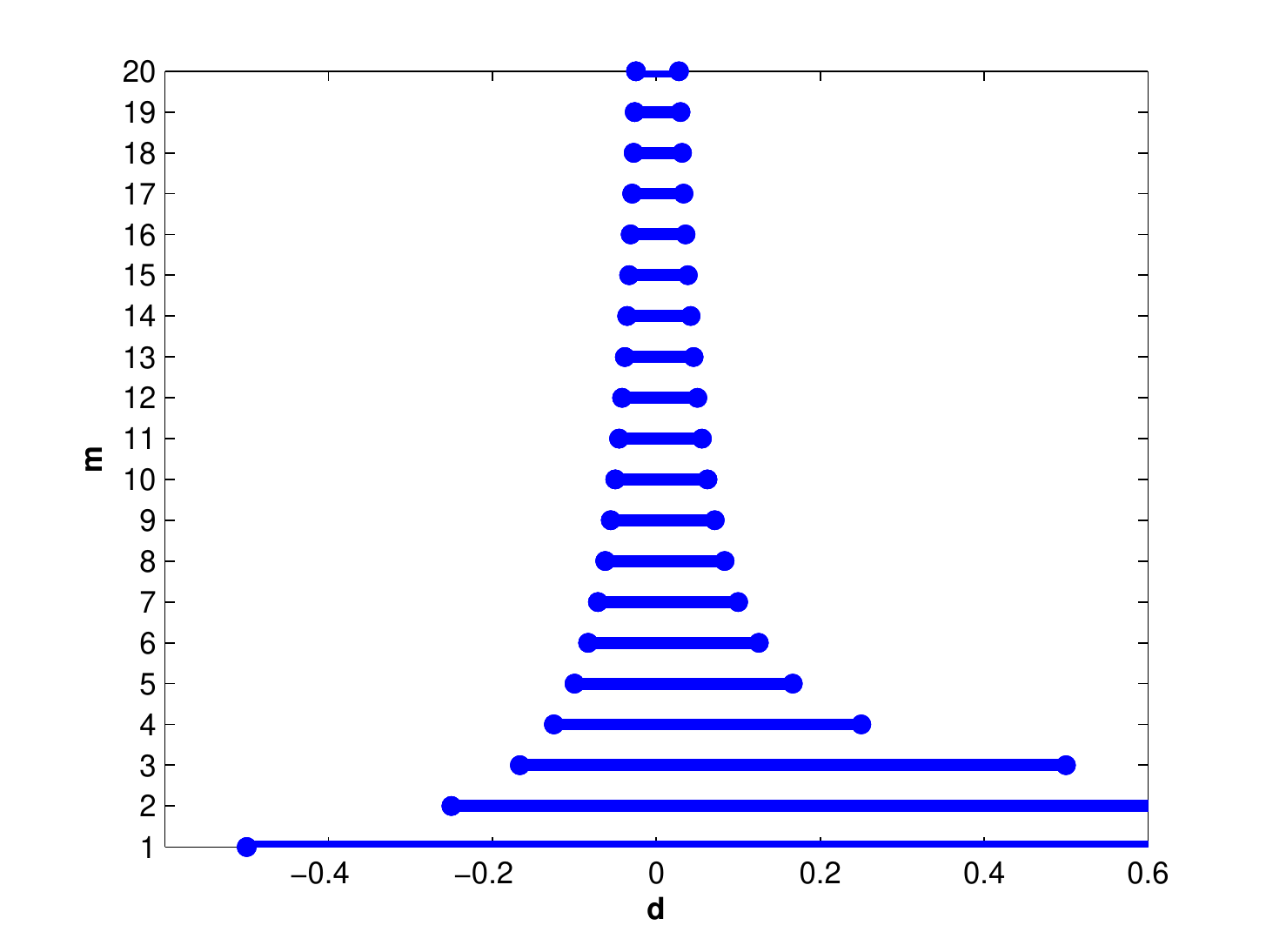}
\par\end{centering}

\protect\caption{Stabililty zone (blue lines) for a duopoly depending of the number
of markets and the values for $d$.\label{fig:Stabililty-zone}}
\end{figure}

\par\end{center}

\section{Numerical simulations}

Using numerical simulations we can see how the complex dynamics of
the equilibrium depends on the scale parameter $\left(d\right)$,
using a scheme of duopolists competing on three markets ($2\times3$
game). The results of our model (blue) will be compared with the base
model (red) proposed by Fisher (see section \ref{Fisher}) under identical
parameter values.

\begin{figure}
\subfloat[$d=-0.2$\label{fig:d=00003D-0.2}]{\protect\includegraphics[width=0.5\textwidth]{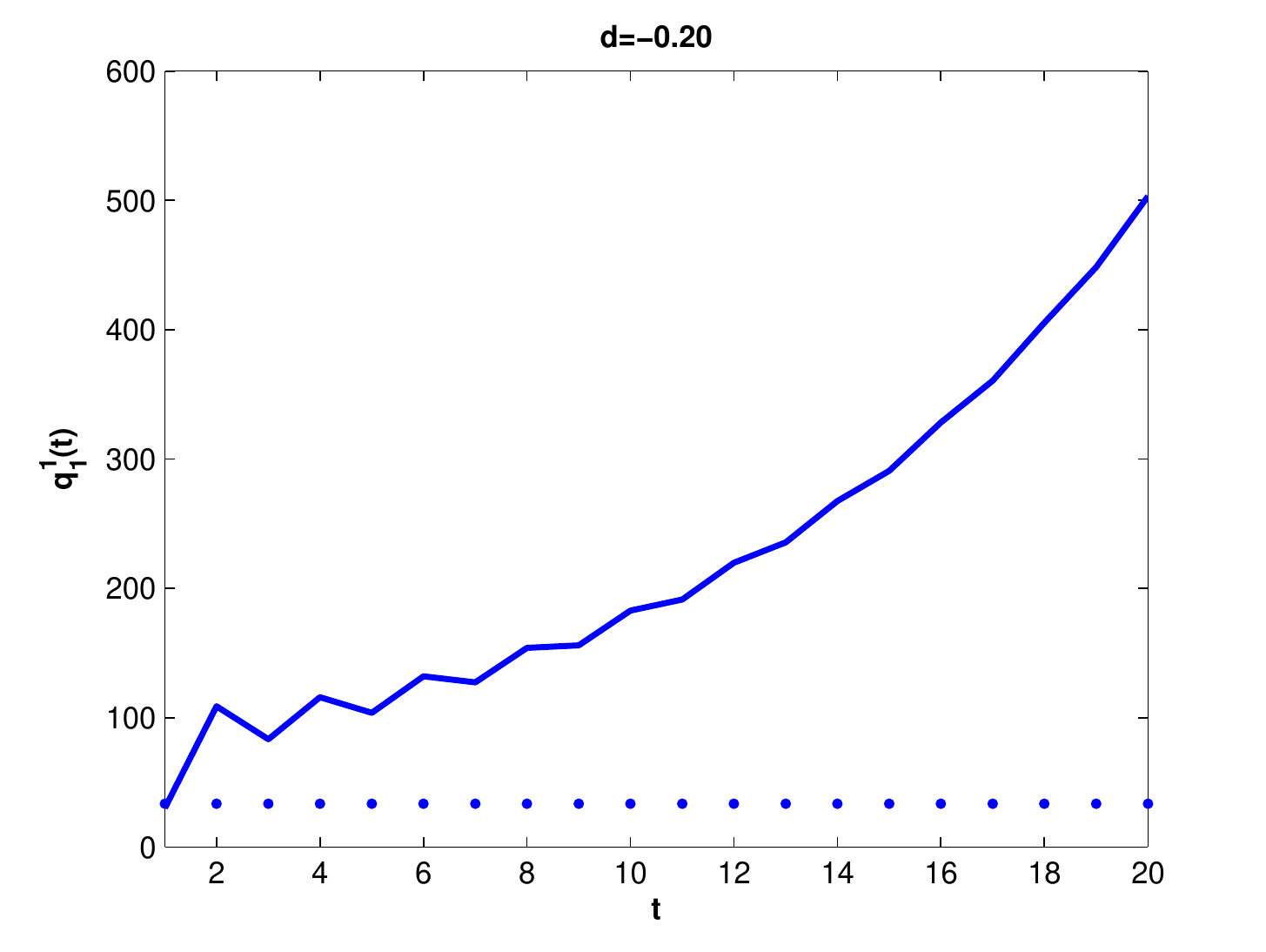}s\protect\includegraphics[width=0.5\textwidth]{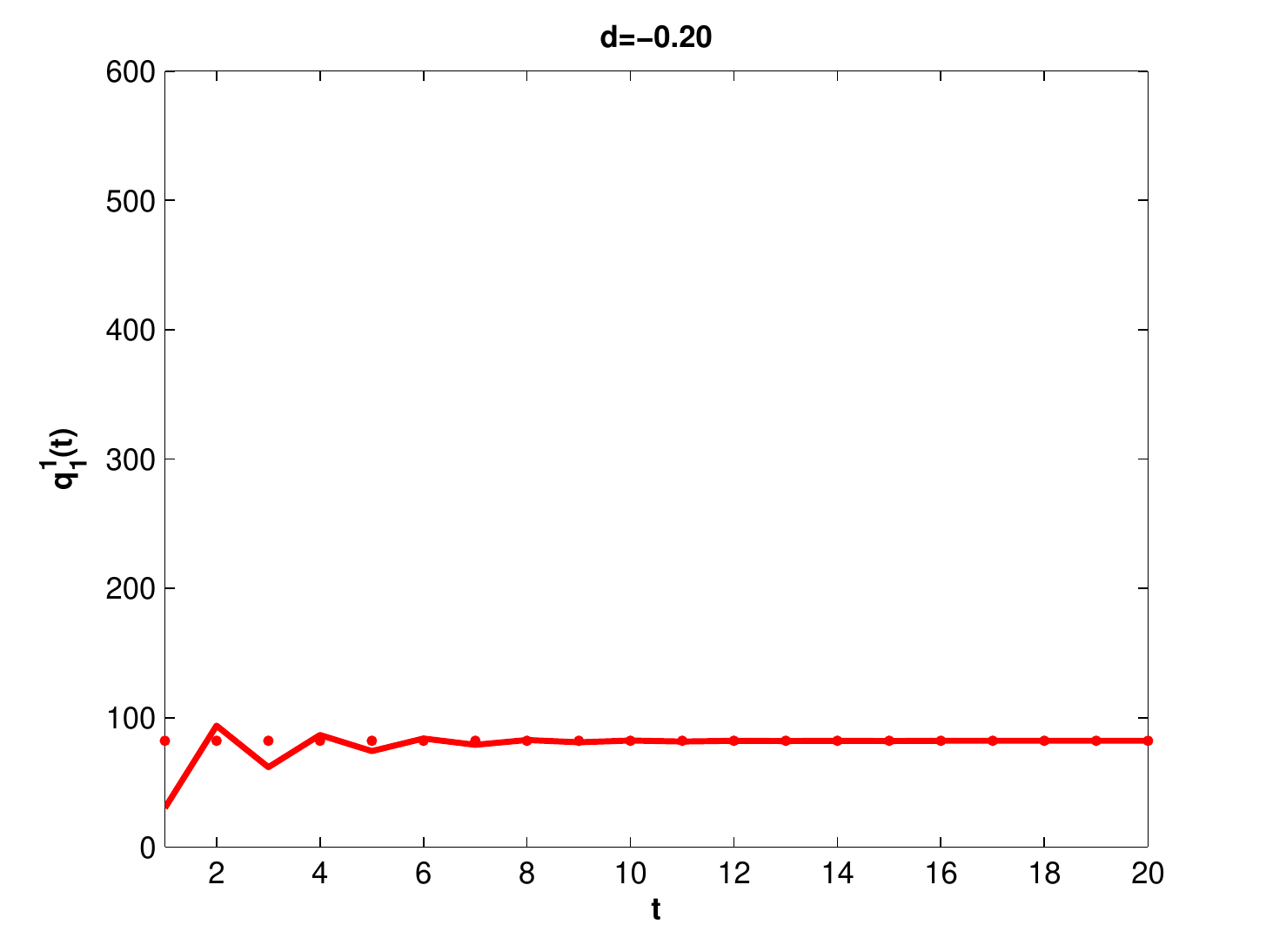}}\\

\subfloat[$d=-0.1$\label{fig: d=00003D0.1}]{\protect\includegraphics[width=0.5\textwidth]{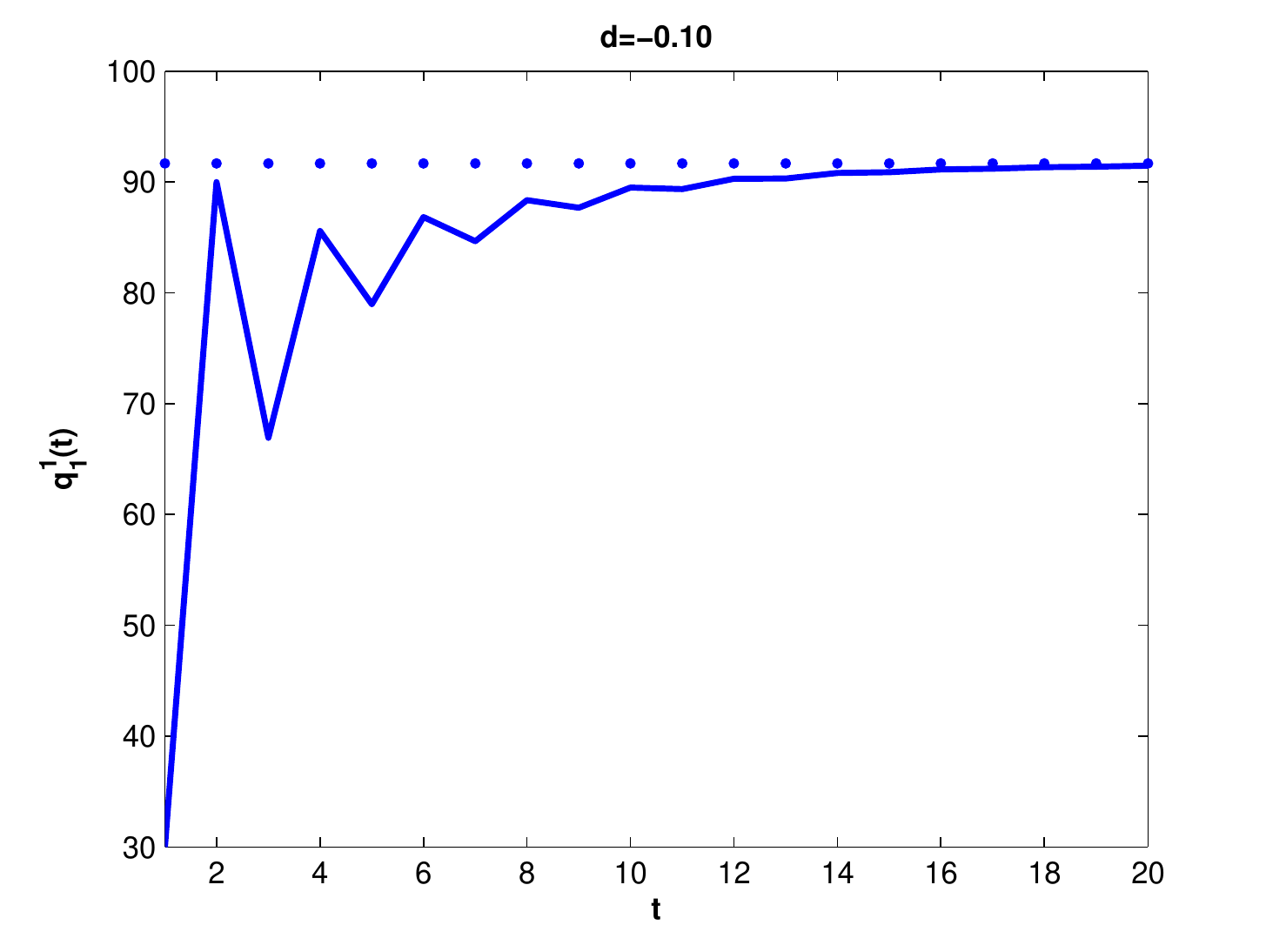}\protect\includegraphics[width=0.5\textwidth]{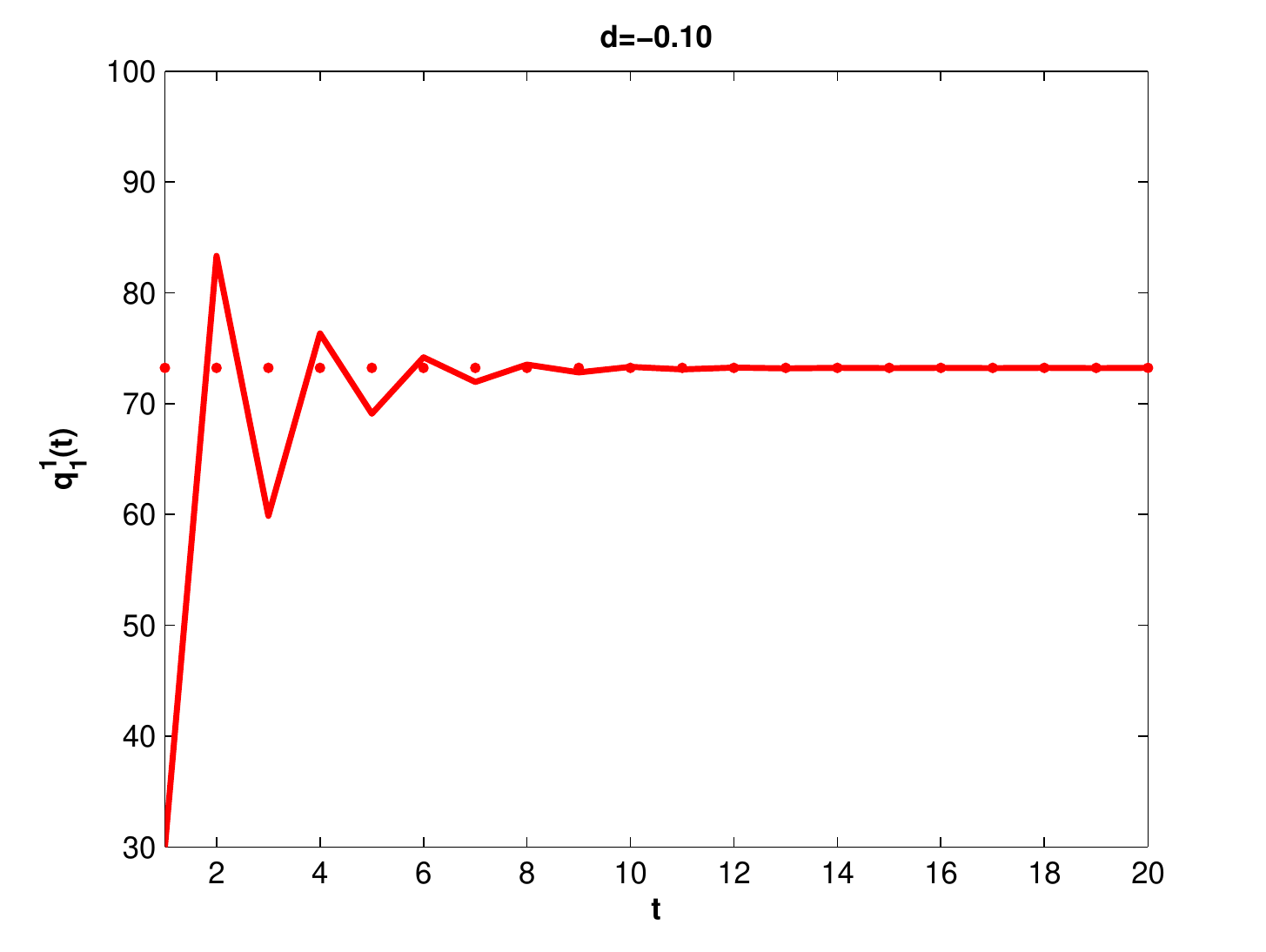}}\\
\subfloat[$d=0$\label{fig:d=00003D0}]{\protect\includegraphics[width=0.5\textwidth]{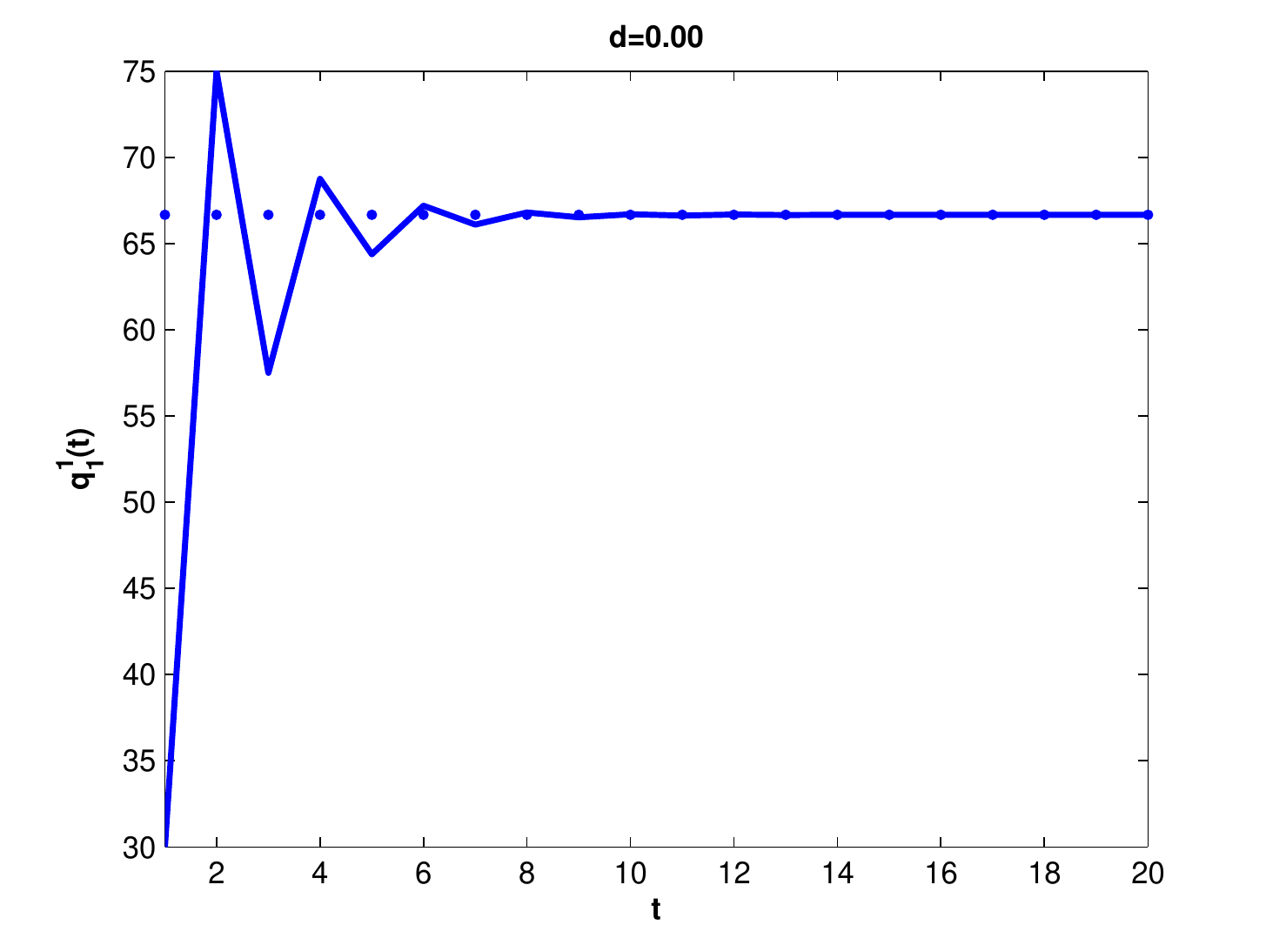}\protect\includegraphics[width=0.5\textwidth]{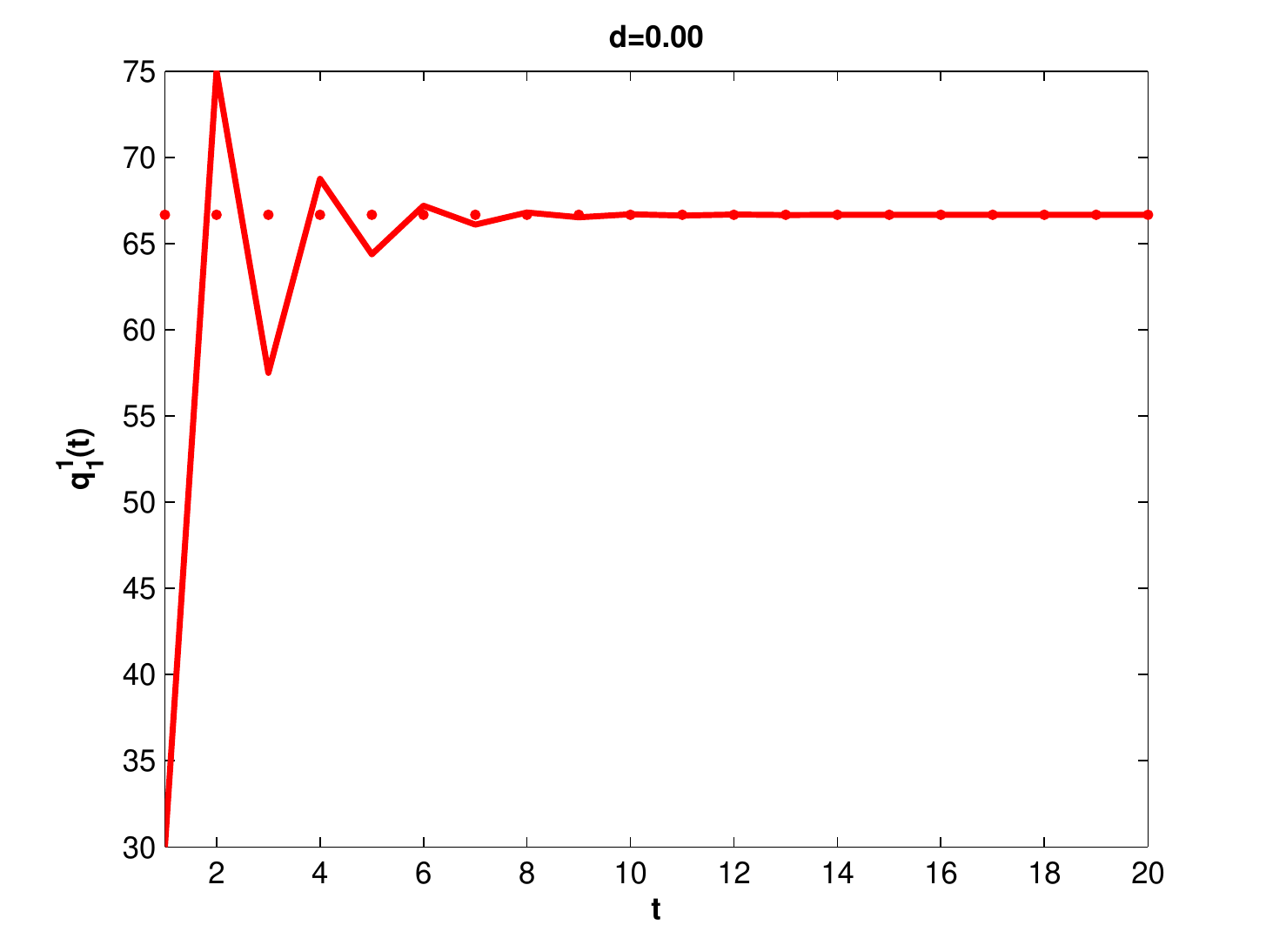}}

\protect\caption{Dynamic of $q_{1}^{1}$ (solid line) and Cuornot equilibrium (dots),
for different values of $d$, under the proposed retail competition
(blue) and the Fisher approach (independent sellers) (red); with $a_{1}=200$,
$a_{2}=150$, $a_{3}=100$, $c_{1}=20$, $c_{2}=40$. \label{fig:Path-duopoly}}
\end{figure}

\begin{figure}
\subfloat[$d=0.2$\label{fig:d=00003D0.2}]{\protect\includegraphics[width=0.5\textwidth]{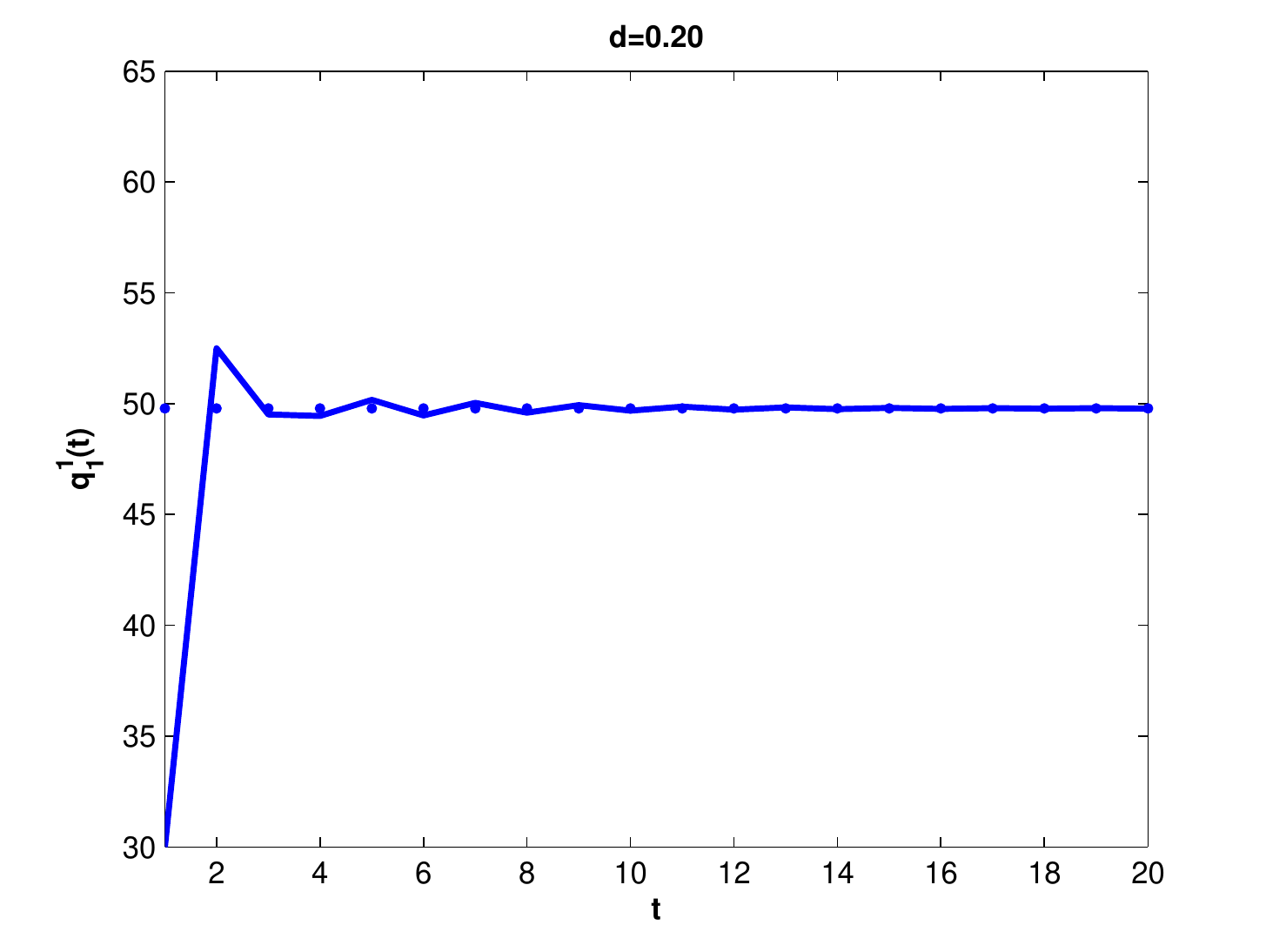}\protect\includegraphics[width=0.5\textwidth]{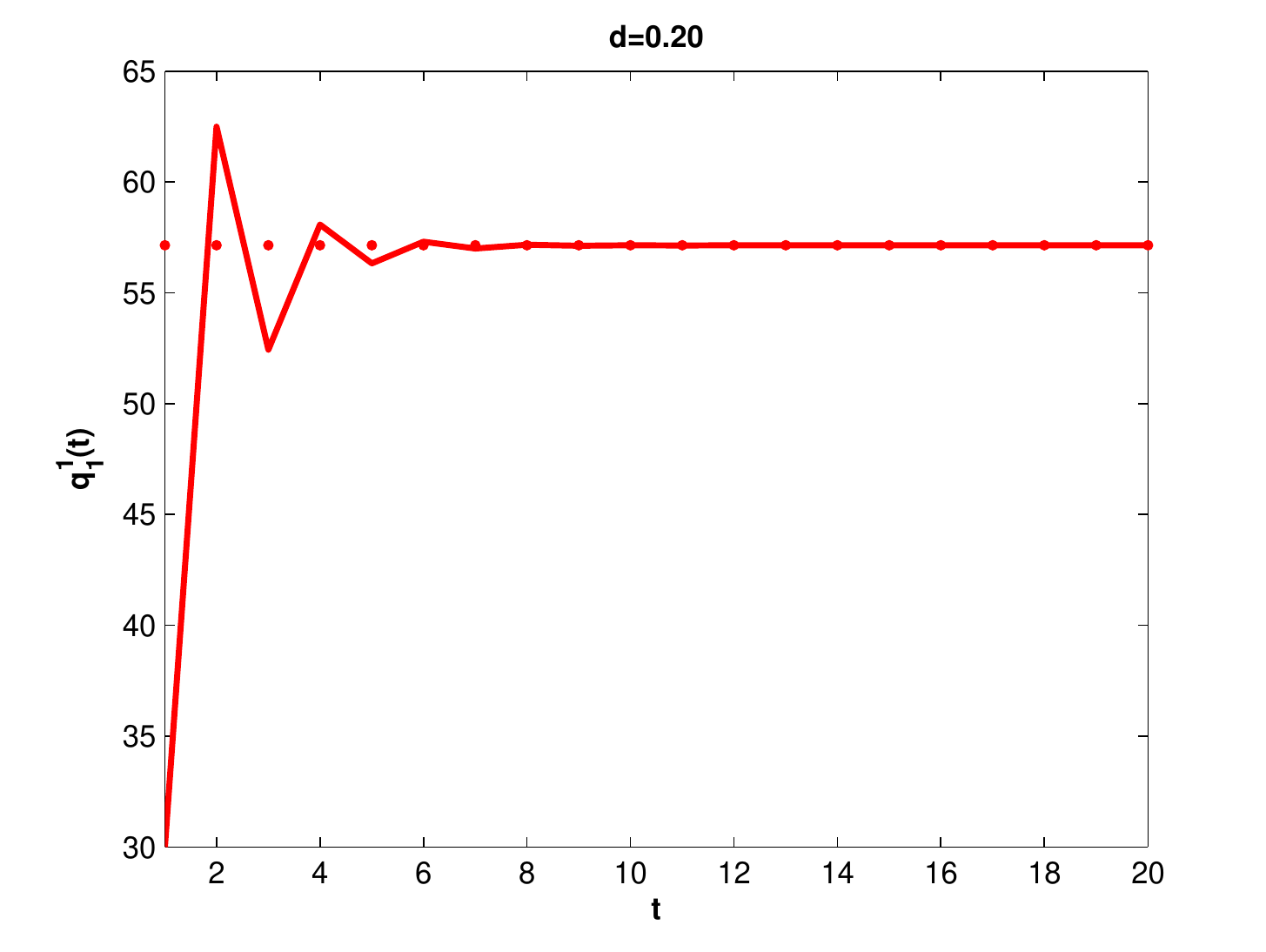}}\\

\subfloat[$d=0.5$\label{fig:d=00003D0.5}]{\protect\includegraphics[width=0.5\textwidth]{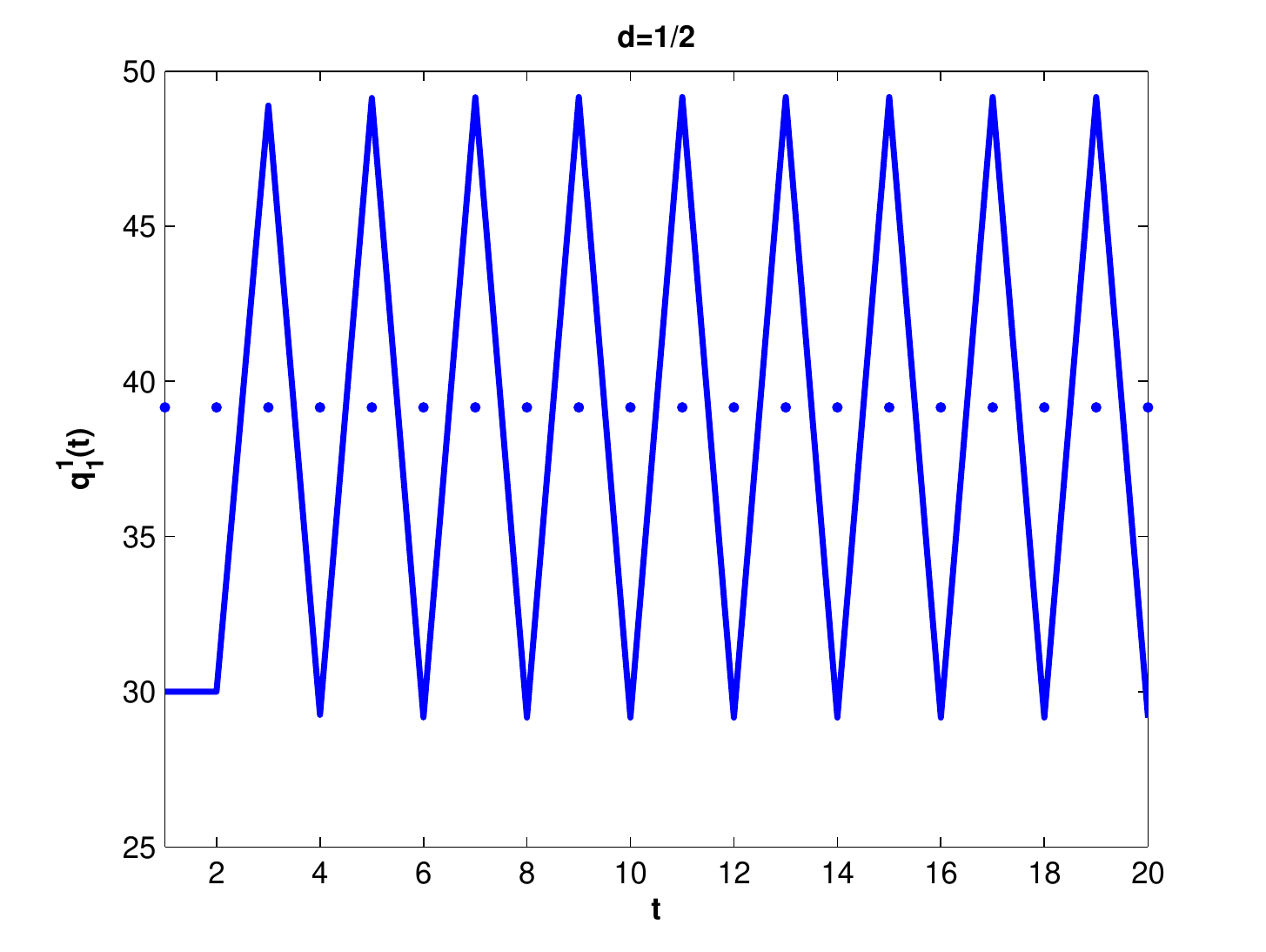}\protect\includegraphics[width=0.5\textwidth]{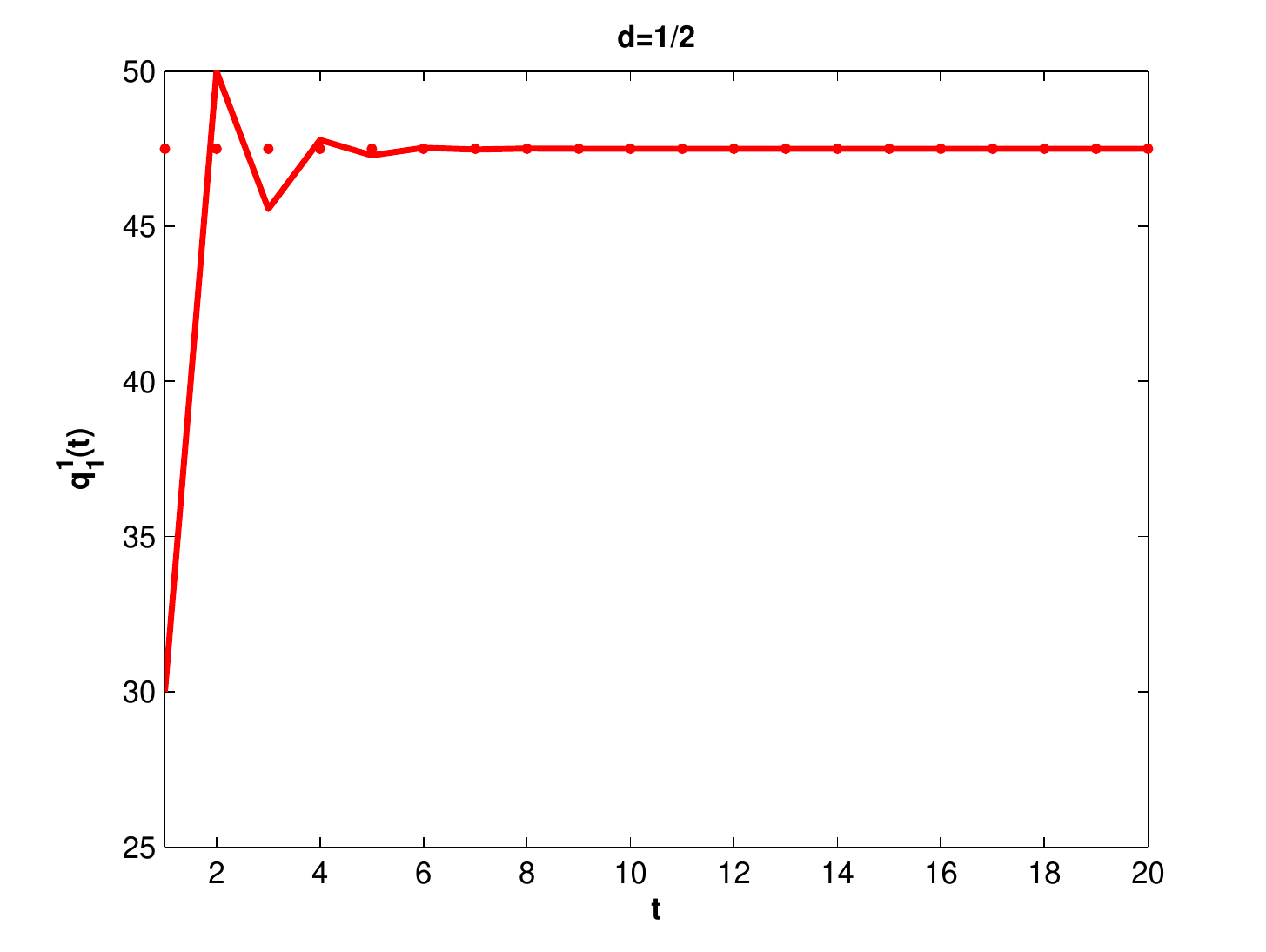}}\\
\subfloat[ $d=0.55$\label{fig:d=00003D.55}]{\protect\includegraphics[width=0.5\textwidth]{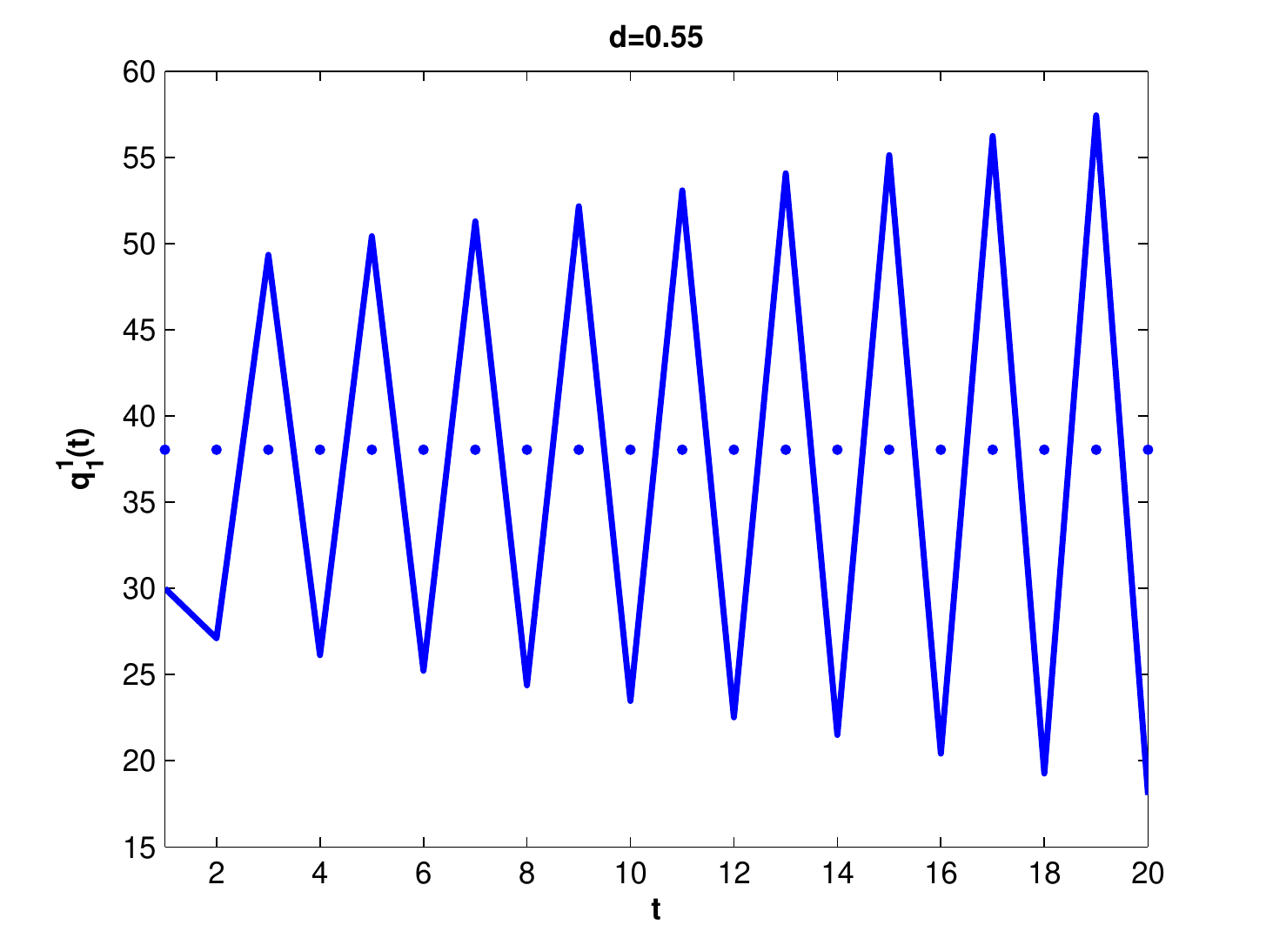}\protect\includegraphics[width=0.5\textwidth]{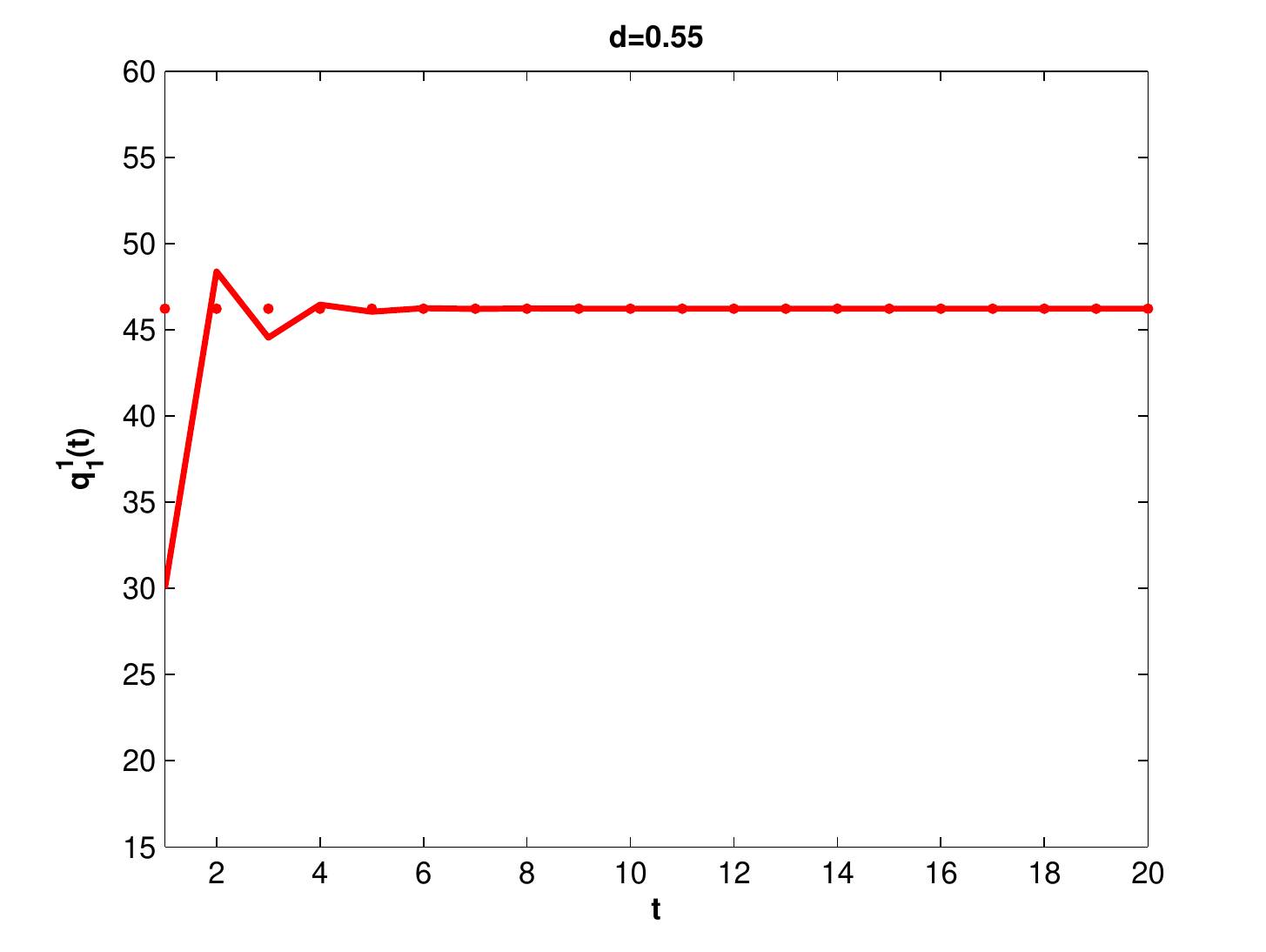}}

\protect\caption{Dynamic of $q_{1}^{1}$ (solid line) and Cuornot equilibrium (dots),
for different values of $d$, under the proposed retail competition
(blue) and the Fisher approach (red); with $a_{1}=200$, $a_{2}=150$,
$a_{3}=100$, $c_{1}=20$, $c_{2}=40$. \label{fig:Path-duopoly-1}}
\end{figure}

In the figs. \ref{fig:Path-duopoly} and \ref{fig:Path-duopoly-1}
we show the dynamic of the quantity allocated by player 1 in market
1 according to several values of $d$. 

For $d=-0.1$ and $d=0.2$ (figs. \ref{fig: d=00003D0.1} and \ref{fig:d=00003D0.2})
we have that the equilibrium under naive expectations reaches the
equilibrium in both models. 

When $d=0$ (fig. \ref{fig:d=00003D0}), the models no longer have
advantages/disadvantages of scale production, being both the same
schemes that the Cournot-Teocharis approach (see section \ref{Fisher}). 

As seen in the fig. \ref{fig:d=00003D0.5}, for the critical point
$d=1/2$, the dynamic goes to stationary oscillations (neutrally stable)
for the retail competition while for the single-market approach arrives
quickly to the equilibrium . 

For the case of figs. \ref{fig:d=00003D-0.2} and \ref{fig:d=00003D.55}
the dynamic of the Fisher model remains stable but in our model lead
to a chaotic behavior. 

\begin{figure}
\begin{centering}
\subfloat[{$d\in[-0.17;-0.16]$}]{\protect\includegraphics[width=0.5\textwidth]{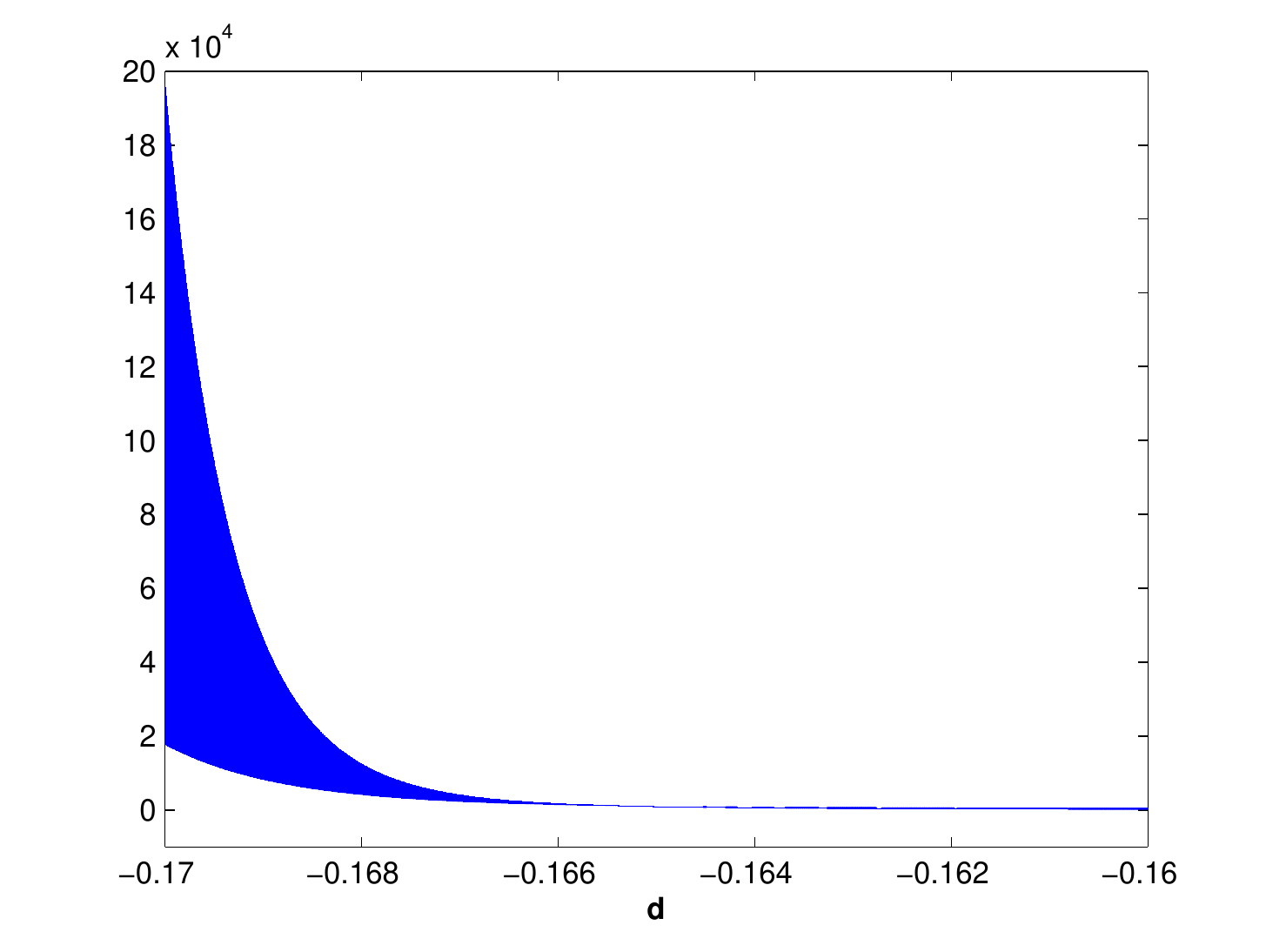}
\protect\includegraphics[width=0.5\textwidth]{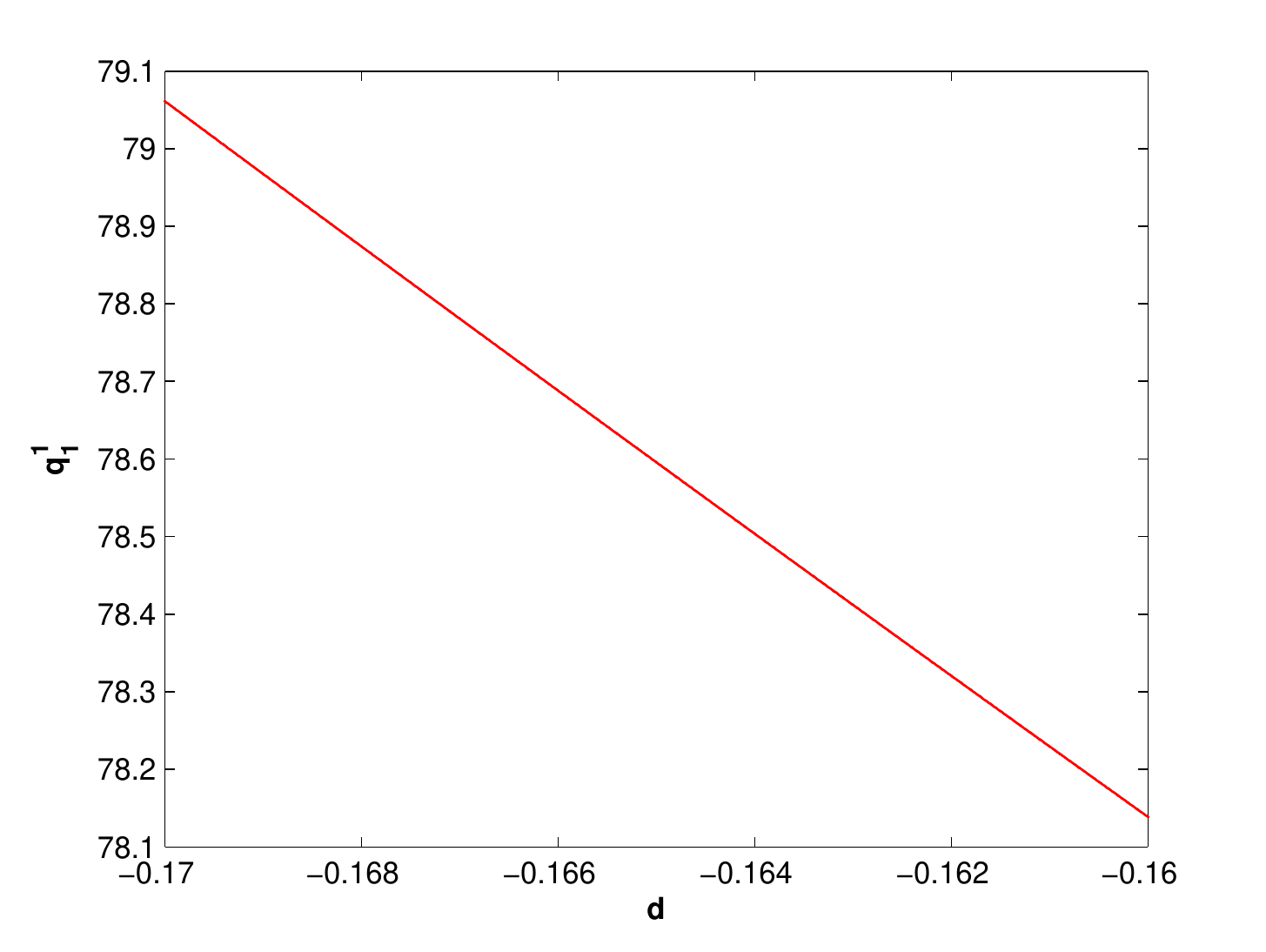}}\\
\subfloat[{$d\in[-0.16;0.45]$\label{fig:sub b}}]{\protect\centering{}\protect\includegraphics[width=0.5\textwidth]{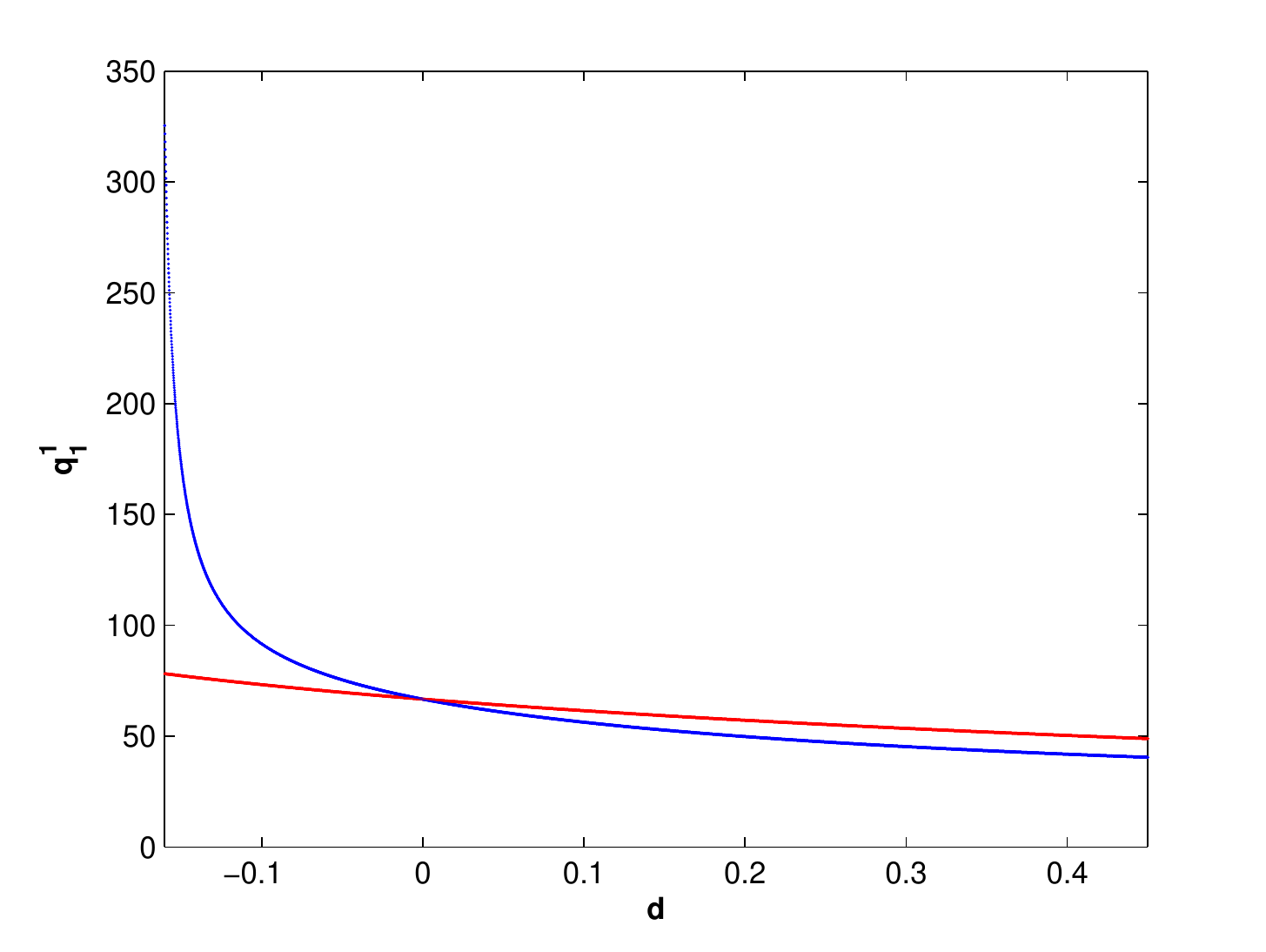}\protect}\\

\par\end{centering}

\subfloat[{$d\in[-0.45;0.52]$}]{\protect\includegraphics[width=0.5\textwidth]{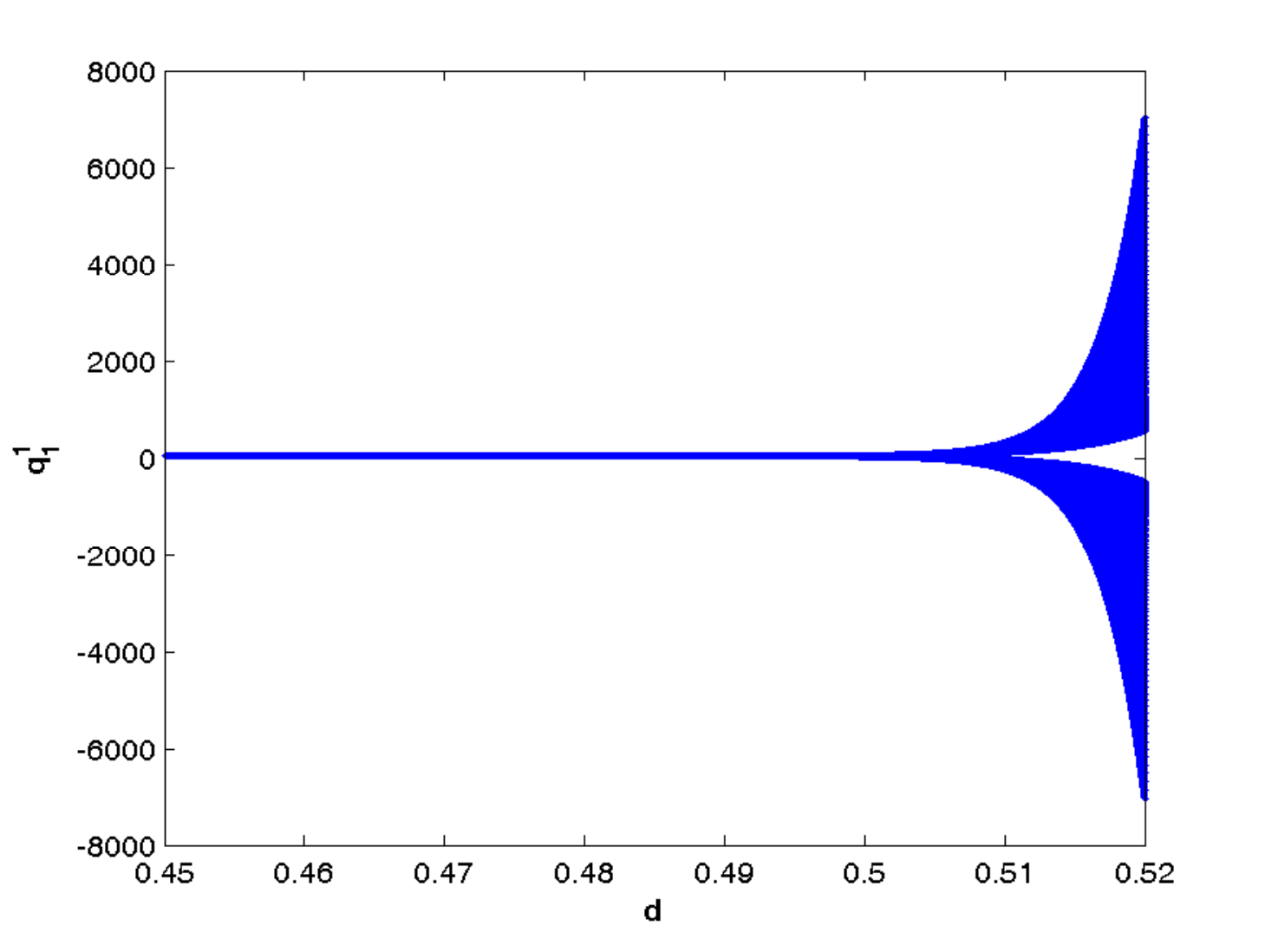}\protect\includegraphics[width=0.5\textwidth]{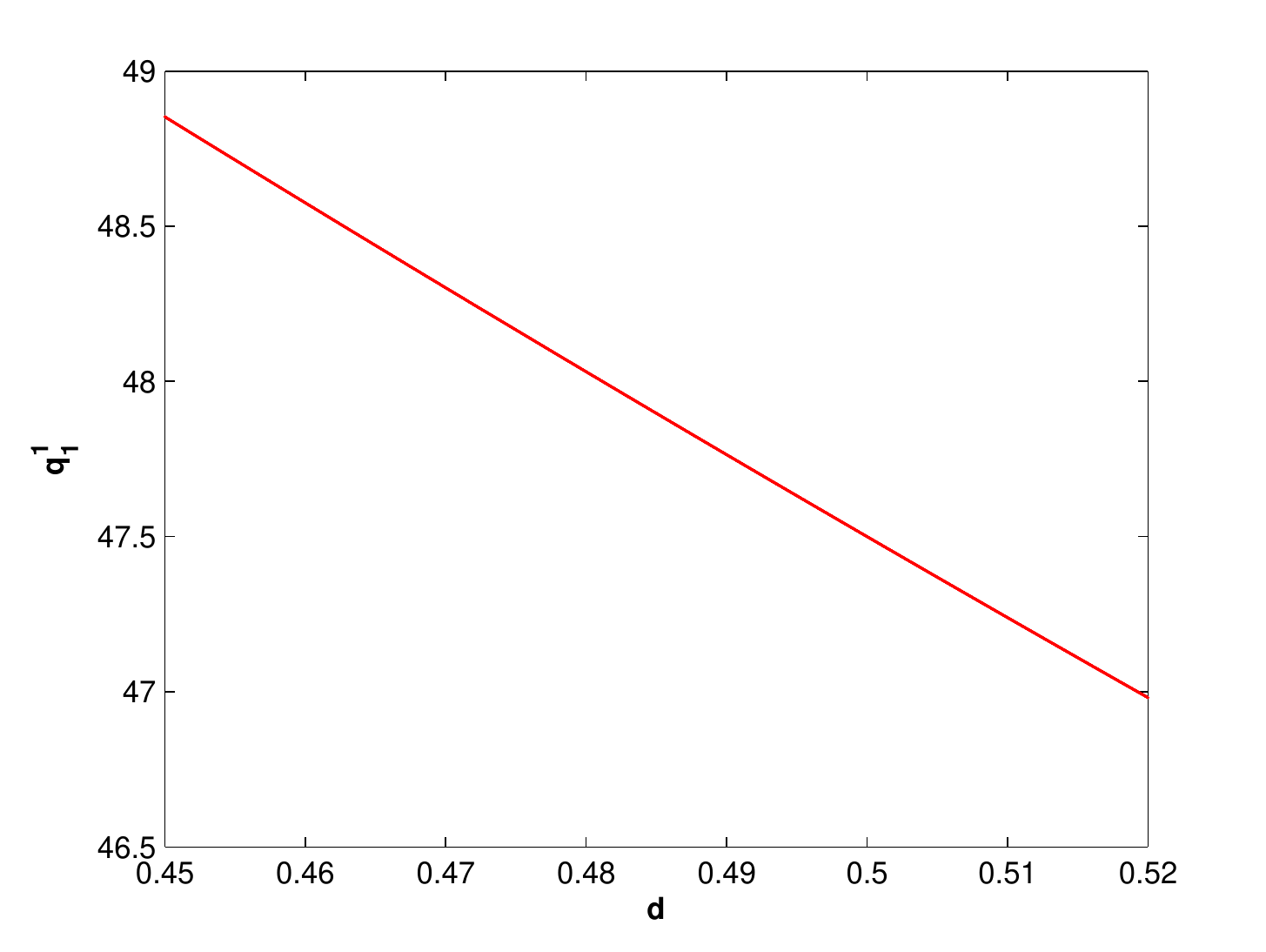}}

\protect\caption{Bifurcation diagrams for the quantity of the player 1 over market
1 using the proposed retail competition model (blue) and the Fisher
approach (red) with $a_{1}=200$, $a_{2}=150$, $a_{3}=100$, $c_{1}=20$
and $c_{2}=40$. \label{fig:Bifurcation-diagrams-zero}}
\end{figure}

The figure \ref{fig:Bifurcation-diagrams-zero} show the complex dynamic
of the the quantity $q_{1}^{1}$ by means of bifurcation diagrams,
using values of $d$ from the interval $[-0.17;0.52]$ . The decentralized
model shows stability of the equilibrium for all the values examined.
By contrast, in our model the chaos is achieved outside the critical
points ($d<-1/6$ and $d>1/2$, as it was predicted by the analytical
results of the duopoly case, see eq. \ref{eq:results duopoly}). In
the stability zone (fig. \ref{fig:sub b}), we can observe how the
centralized managerial decision takes advantage in terms of production
when the company operates under economies of scale in relation to
the disaggregated model; while at diseconomies of scale, the production
performance of the local model are less affected than the multi-market
scheme.\\

\section{Conclusions}

In this work, we analyze the stability of a multi-store retail competition
model. Specifically, we model an oligopoly system with multi-market
competition. The model considers the impact of the firm's economies
or diseconomies of scale. In one extreme, the multi-store retail maintain
a centralized decision making process, optimizing global economies
of scale due to its global size. On the other hand, we have local
oligopolistic competition, where the same demand is served by different
firms that compete in only one market. Our model confirms the fact
that economies and diseconomies of scale make the Cournot equilibrium
very unstable for certain values of the scale parameter of the producers.
Additionally, the number of markets, as expected, tends to contribute
to this instability. One very interesting further research is to expand
the multi-market oligopolistic model to a multi-product setting.\\

\section{Acknowledgments}

Aid from the Fondecyt Program, project Nº 1131096, is grateful acknowledged
by Marcelo Villena. Axel Araneda thanks as well Faculty of Engineering
\& Science of the Universidad Adolfo Ib\'a\~nez for financial support.\\

\bibliographystyle{unsrt}
\bibliography{18C__Users_Achel_UAI_DISC_Research_Stability_Stability}

\end{document}